\documentclass[letterpaper]{article} % DO NOT CHANGE THIS
\usepackage{aaai24}  % DO NOT CHANGE THIS
\usepackage{times}  % DO NOT CHANGE THIS
\usepackage{helvet}  % DO NOT CHANGE THIS
\usepackage{courier}  % DO NOT CHANGE THIS
\usepackage[hyphens]{url}  % DO NOT CHANGE THIS
\usepackage{graphicx} % DO NOT CHANGE THIS
\usepackage{natbib}  % DO NOT CHANGE THIS AND DO NOT ADD ANY OPTIONS TO IT
\usepackage{caption} % DO NOT CHANGE THIS AND DO NOT ADD ANY OPTIONS TO IT
\usepackage{algorithm}
\usepackage{algorithmic}

\usepackage{amssymb}
\usepackage{amsmath}
\usepackage{graphicx}
\usepackage{float}
\usepackage{subfigure}
\usepackage{makecell}
\usepackage{multirow}
\usepackage[table,xcdraw]{xcolor}
\usepackage{enumitem}

\usepackage{newfloat}
\usepackage{listings}
\DeclareCaptionStyle{ruled}{labelfont=normalfont,labelsep=colon,strut=off} % DO NOT CHANGE THIS
\floatstyle{ruled}
\newfloat{listing}{tb}{lst}{}
\floatname{listing}{Listing}
\title{Frequency Spectrum is More Effective for Multimodal Representation and Fusion: A Multimodal Spectrum Rumor Detector}
\author{
    An Lao\textsuperscript{\rm 1}, Qi Zhang\textsuperscript{\rm 2,4},
    Chongyang Shi\textsuperscript{\rm 1}\thanks{Corresponding authors}, 
    Longbing Cao\textsuperscript{\rm 3},
    Kun Yi\textsuperscript{\rm 1},
    Liang Hu\textsuperscript{\rm 2,4},
    Duoqian Miao\textsuperscript{\rm 2}\\
}
\affiliations{
    \textsuperscript{\rm 1}Beijing Institute of Technology, 
    \textsuperscript{\rm 2}Tongji University,
    \textsuperscript{\rm 3}Macquarie University,
    \textsuperscript{\rm 4}DeepBlue Academy of Sciences\\
    \{an.lao, cy\_shi,yikun\}@bit.edu.cn, \{zhangqi\_cs, lianghu, dqmiao\}@tongji.edu.cn,\ longbing.cao@mq.edu.au
}

\usepackage{bibentry}
\begin{document}

\maketitle

\begin{abstract}
Multimodal content, such as mixing text with images, presents significant challenges to rumor detection in social media. Existing multimodal rumor detection has focused on mixing tokens among spatial and sequential locations for unimodal representation or fusing clues of rumor veracity across modalities. However, they suffer from less discriminative unimodal representation and are vulnerable to intricate location dependencies in the time-consuming fusion of spatial and sequential tokens. This work makes the first attempt at multimodal rumor detection in the frequency domain, which efficiently transforms spatial features into the frequency spectrum and obtains highly discriminative spectrum features for multimodal representation and fusion. A novel \textbf{F}requency \textbf{S}pectrum \textbf{R}epresentation and f\textbf{U}sion network (FSRU) with dual contrastive learning reveals the frequency spectrum is more effective for multimodal representation and fusion, extracting the informative components for rumor detection. FSRU involves three novel mechanisms: utilizing the Fourier transform to convert features in the spatial domain to the frequency domain, the unimodal spectrum compression, and the cross-modal spectrum co-selection module in the frequency domain. Substantial experiments show that FSRU achieves satisfactory multimodal rumor detection performance.
\end{abstract}

\section{Introduction}

With the rapid development of social media in various aspects of our lives, the prevalence of content from multiple sources and in diverse formats has significantly increased. A prime example is the combination of text of varying lengths accompanied by images. However, along with this proliferation of multimodal media, a more sophisticated and concerning issue has arisen: multimodal rumors. Multimodal rumors refer to disseminating misinformation or false information through social media platforms, incorporating multiple modes of communication such as text and images. These rumors often defy logical reasoning and lack credibility. Research reveals that rumors are shared more extensively on Facebook than on mainstream news~\cite{willmore2016analysis}. As a result, it has become imperative to detect and mitigate multimodal rumors to effectively manage the associated risks and ensure compliance with social media norms and guidelines \cite{allcott2017social,10130819}.

Recent studies of multimodal rumor detection primarily focus on two key aspects: learning spatial and sequential dependencies in uni-modality and fusing evidence of rumor veracity across different modalities \cite{LZSLTS22, ZhengZGWZ022, SinghalPMSK22}. 1) To obtain informative uni-modal representation, researchers have employed various neural models, such as Convolutional Neural Networks (CNNs), Recurrent Neural Networks (RNNs), and Transformers to perform token mixing over spatial locations of images or sequential positions of text. However, these methods suffer from less discriminative unimodal representation, hindering subsequent fine-grained cross-modal fusion. 2) Existing approaches often apply contrastive learning~\cite{ying2023bootstrapping} or co-attention mechanisms~\cite{QianWHFX21} to achieve multimodal alignment or fusion for detecting rumor across modalities. However, they may either overlook the interpretable fine-grained fusion or encounter intricate location dependencies in fusing spatial and sequential tokens. Moreover, current approaches for fine-grained fusion, such as co-attention mechanisms, often exhibit quadratic time complexity \cite{RaoZZLZ21}. These issues collectively undermine the accuracy and efficiency of multimodal rumor detection models, highlighting the need for further advancements in this field.

\begin{figure}[htbp]
    \centering
    \includegraphics[width=0.47\textwidth]{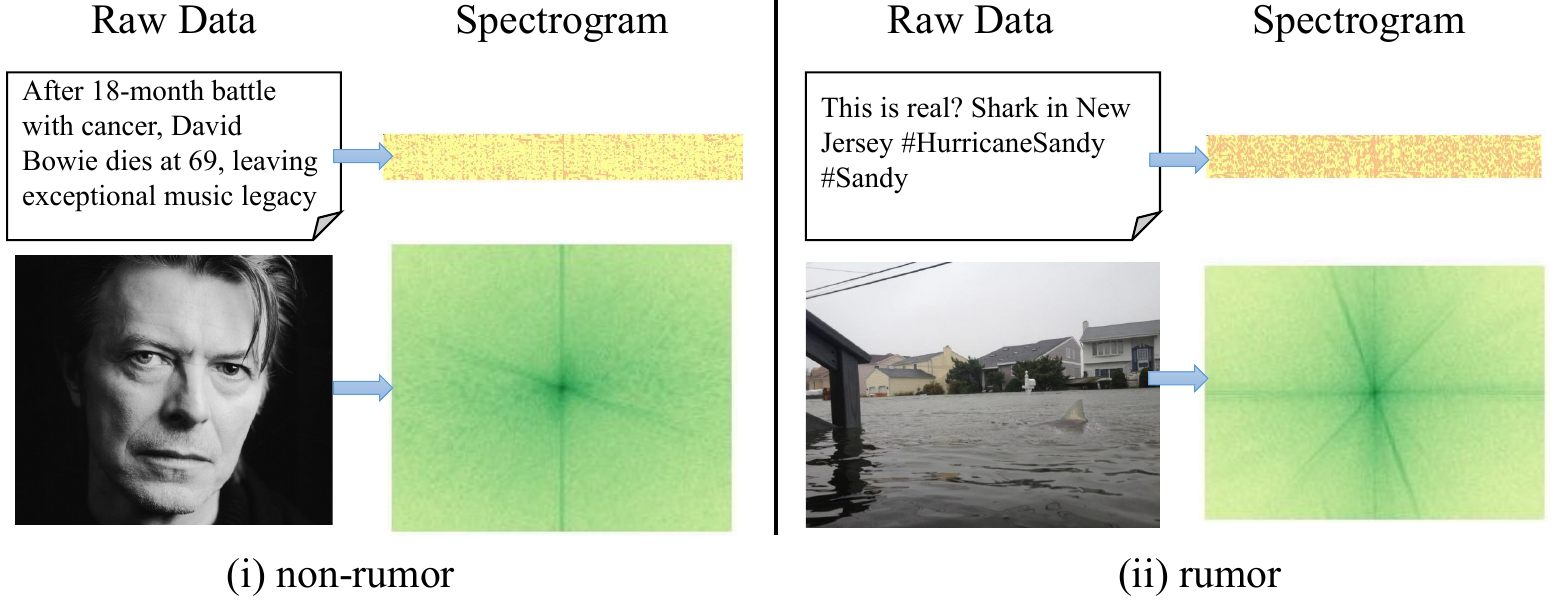}
    \vspace{-5mm}
    \caption{Two examples on Twitter visualize the raw data and its spectrograms. It shows the spectrum has centrally concentrated components and discriminative patterns.}
    \label{fig:problem}
    \vspace{-4mm}
\end{figure}

To address the issues, we make the first attempt from a new paradigm and architecture in this work: multimodal spectrum rumor detection. We contend that the frequency spectrum offers a more effective means of representing and fusing multimodal data. Inspired by signal processing theories~\cite{MateosSMR19}, we can utilize Fourier transforms to transform sequential (text) or spatial (images) data to the frequency domain. The Fourier transform often generates \textit{a sparse frequency spectrum} with a significant portion of frequency components approaching zero (shown in Figure \ref{fig:problem}). This characteristic facilitates obtaining discriminative uni-modal representation and emphasizing (suppressing) veracity-relevant (irrelevant) features for detection. In addition, the frequency spectrum provides \textit{a global view}~\cite{RaoZZLZ21}, allowing each spectrum component to attend to all features in the spatial domain. Unlike the position-based alignment in co-attention mechanisms~\cite{ZhengZGWZ022}, the spectrum exhibits global patterns (see Figure~\ref{fig:problem}), allowing a more comprehensive sense of intricate location dependencies within/across modalities between rumors and non-rumors. Moreover, point-wise multiplication in the frequency domain is equivalent to self-attention in the spatial domain, avoiding quadratic time complexity (Appendix A).

Accordingly, we propose an architecturally simple and computationally efficient multimodal spectrum rumor detector: a Frequency Spectrum Representation and fUsion network (FSRU) with dual contrastive learning. FSRU comprises three key components: text and image embedding, multimodal frequency spectrum representation and fusion module, and detection with distribution similarity.
Especially, the frequency spectrum representation and fusion module includes four core operations: we introduce
1) discrete Fourier transform (DFT) to convert features in the spatial domain to the frequency domain;
% an elemental multiplication between the uni-modal frequency compression and uni-modal features, 
2) unimodal spectrum compression to compress frequency domain features; 
3) cross-modal spectrum co-selection to select spectrum components; 
and 4) inverse DFT (IDFT) to reverse frequency domain features to the spatial domain. 
By utilizing filter banks in the frequency domain, unimodal spectrum compression generates spectral compressed representations to reveal potential features within each modality and portray distinct feature patterns. 
Cross-modal spectrum co-selection makes use of complementary dependencies between modalities to select informative spectrum components that are beneficial in identifying rumors. 
Subsequently, we devise a fusion module that leverages the similarity of feature distributions to generate a cohesive multimodal representation and introduce dual contrastive learning to enhance multimodal learning. We conduct experiments on two real-world datasets to evaluate our proposed approach, FSRU. The results demonstrate that FSRU yields favorable outcomes across different evaluation metrics and aspects.

Our contributions are twofold:
\begin{itemize}
    \item An architecturally simple and computationally efficient novel method \textbf{F}requency \textbf{S}pectrum \textbf{R}epresentation and f\textbf{U}sion network (FSRU) with dual contrastive learning is proposed for multimodal rumor detection. Unlike existing approaches that primarily focus on features in the spatial/sequential domain, FSRU aims to capture discriminative unimodal features and fuse cross-modal evidence of rumor veracity in the frequency domain. This architecturally simple approach offers a fresh perspective on multimodal rumor detection.
    \item A frequency spectrum representation and fusion module is proposed to extract rumor evidence that is concealed in the frequency components from both unimodal and cross-modal perspectives. The unimodal spectrum compression explores clearer patterns in text and image representations. The cross-modal spectrum co-selection guides retaining relevant frequency components while fusing multimodal spectrum features, effectively reducing the impact of irrelevant frequency components.
\end{itemize}

\section{Related work}
\subsection{Multimodal Rumor Detection}
Previous work attempts to solve multimodal rumor detection by concatenating text and image features \cite{WangMJYXJSG18, CuiW019, SinghalS0KS19, ZhangWCZGMC20}. 
They concatenate multimodal features from the spatial dimension without considering modal interactions. 
To address this deficiency, MFN \cite{ChenWY0WL21} employs a self-attentive fusion module to capture the relationships between text and image.
CAFE \cite{LZSLTS22} introduces cross-modal alignment and ambiguity learning to learn cross-modal correlations while integrating multimodal features.
% Both methods characterize text and image features separately.
% In addition, hidden state contextual information complements the modal representation during the feature representation phase, which allows a fine-grained extraction of intrinsic correlations between modalities and accurate learning of rumor-related features.
Hidden state contextual information complements the modal representation during the feature representation phase. 
Sun et al. \cite{SunZML21} design a modality-shared embedding and introduce external knowledge to assist with rumor detection.
% MM-MTL \cite{ZhangQFX22} presents a feature-sharing multi-task learning method to generate precise multimodal representations. 
% Other studies \cite{ZhangFQX19, SunZML21, LiQLZ22, LiQLZ22GCN}  capture feature inconsistencies by introducing external knowledge to indicate the truth of rumors and improve modeling performance. 
Recently, attention-based functions have been popularly involved in multimodal rumor detection.
MFAN \cite{ZhengZGWZ022} enhances the model representation by extracting mutual information between modalities through cross-modal co-attention mechanisms.
To improve the multimodal learning capability, HMCAN \cite{QianWHFX21} adopts a Transformer-based contextual attention network to extract multimodal contextual complementary information. 
BMR \cite{ying2023bootstrapping} proposes the Improved Multi-gate Mixture-of-Expert networks to learn information from unimodal and multimodal features through single-view prediction and cross-modal consistency learning.

\subsection{Fourier Transform in Deep Learning}
Fourier transform plays a vital role in the area of digital signal processing. 
It has been introduced to deep learning for enhanced learning performance \cite{ehrlich2019deep, chi2020fast,li2020falcon,yang2020fda,abs-2302-02173,abs-2311-06190}. 
GFNet \cite{RaoZZLZ21} utilizes fast Fourier transform to convert images to the frequency domain and exchange global information between learnable filters. 
As a continuous global convolution independent of input resolution, Guibas et al. \cite{AFNO} design the adaptive Fourier neural operator frame token mixing. 
Xu et al. \cite{Xu0007QSWCR20} devise a learning-based frequency selection method to identify trivial frequency components and improve the accuracy of classifying images. 
On text classification, Lee-Thorp et al. \cite{eckstein2022fnet} use the Fourier transform as a text token mixing mechanism. 
Furthermore, the Fourier transform is also applied to forecast time series \cite{CaoWDZZHTXBTZ20, LangeBK21, KocK22, YangH22a}. 
To increase the accuracy of multivariate time-series forecasting, Cao et al. \cite{CaoWDZZHTXBTZ20} propose a spectral temporal graph neural network (StemGNN), which mines the correlations and time dependencies between sequences in the spectral domain. 
Yang et al. \cite{YangH22a} propose bilinear temporal spectral fusion (BTSF), which updates the feature representation in a fused manner by explicitly encoding time-frequency pairs and using two aggregation modules: spectrum-to-time and time-to-spectrum. 

Our work is inspired by \cite{RaoZZLZ21, Xu0007QSWCR20, abs-2311-06184} but differs from them. 
% We extend the learnable filter to prune banks to mine more comprehensive and clearer feature patterns via different frequency responses. 
% We also introduce the discrete cosine transform to improve frequency-domain energy aggregation. 
% Furthermore, to the best of our knowledge, we make the first attempt to apply the Fourier transform to both texts and images for multimodal rumor detection. Our work learns the interactions between these multimodal features in the frequency domain. 
To our knowledge, there are no existing techniques for multimodal rumor detection that employ the same architecture for frequency domain characterization as our approach.
Our approach differs from other spatial domain techniques in that we not only convert the original features into the frequency domain but also perform a series of complex-valued computation operations in the frequency domain.

\section{Problem definition}
We formulate multimodal rumor detection as a binary classification task, where multimodal $a$ refers to text and image modalities, denoted as $a\in\{t,v\}$. 
Given a multimodal rumor dataset $\mathcal{D}=\{\mathcal{X},\mathcal{Y}\}$, each sample is denoted as $(x,y)$, and $x$ can be represented by $x=\{x^t, x^v\}$, where $x^t$ stands for text and $x^v$ for image. 
$y\in\{0,1\}$ is the rumor veracity label corresponding to  sample $x$, $y=1$ indicates that the sample is a rumor, while $y=0$ indicates that the sample is true. 
This work aims to incorporate text and image features to predict the rumor label $\hat{y}\in\{0,1\}$. 

\begin{figure*}[htbp]
    \centering
    \includegraphics[width=0.95\textwidth]{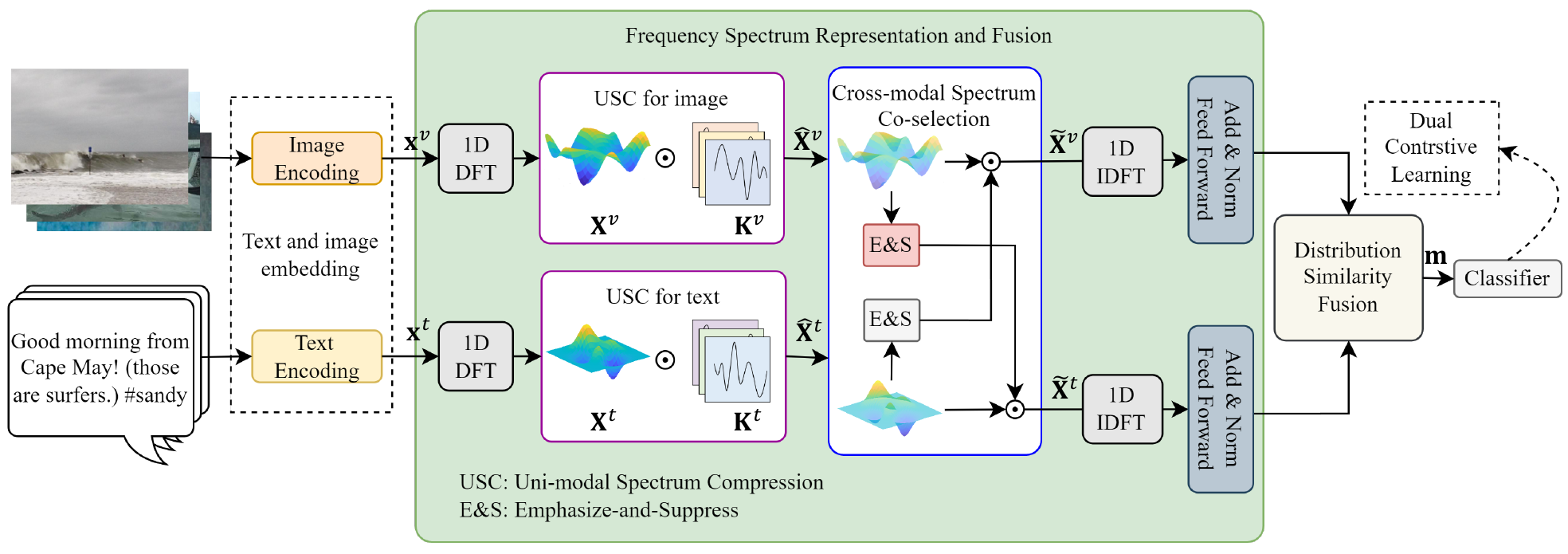}
    \vspace{-2mm}
    \caption{The architecture of our proposed Frequency Spectrum Representation and fUsion network (FSRU) for multimodal rumor detection. FSRU comprises three main components: a text and image embedding module, a frequency spectrum representation and fusion module, and a classification with distribution similarity. }
    \label{fig:model}
    \vspace{-3mm}
\end{figure*}

\section{Methodology}
We propose a Frequency Spectrum Representation and fUsion network (FSRU) with dual contrastive learning to tackle the problem of multimodal rumor detection. 
As illustrated in Figure \ref{fig:model}, FSRU comprises three components:
1) \textit{text and image embedding} obtains textual and visual unimodal embeddings for social media posts through two embedding modules, respectively. 
2) \textit{frequency spectrum representation and fusion module} explores unimodal spectrum information and cross-modal spectrum interactions. 
3) \textit{detection with distribution similarity} performs the final detection after obtaining a multimodal representation by capturing the complementary semantic relationships between unimodality. 
% 4) \textit{dual contrastive learning module}, which reinforces the multimodal representation and improves the robustness of the model. 
Next, we explain each component in detail.

\subsection{Text and Image Embedding}
Given a rumor sample $x=\{x^t, x^v\}$, we first embed its raw text and image, respectively. Regarding the text sequence $x^t=[w_1,w_2,...,w_m]$ ($m$ is the number of words), we simultaneously employ word embedding and positional embedding to encode each word, denoted by: 
\begin{equation}
    \mathbf{w}_i=\operatorname{WE}(w_i)+\operatorname{PE}^t(w_i)
\end{equation}
where $\operatorname{WE}(\cdot)$ is the word embedding and $\operatorname{PE}^t(\cdot)$ is the position embedding for text sequence. Accordingly, we obtain the text embedding $\mathbf{x}^t=[\mathbf{w}_1,\mathbf{w}_2,...,\mathbf{w}_m]$. Regarding images, we divide each image into $h\times w$ non-overlapping patches 
$x^v=[p_1,p_2,...,p_n]$ ($n=h\times w$) and adopt CNN \cite{lecun1995convolutional} to generate meaningful representations:
\begin{equation}
    \mathbf{p}_i=\operatorname{CNN}(\operatorname{PE}^v(p_i))
\end{equation}
where $\operatorname{PE}^v(\cdot)$ is the patch embedding for the image. We obtain the image embedding $\mathbf{x}^v=[\mathbf{p}_1,\mathbf{p}_2,...,\mathbf{p}_n]$.

\subsection{Frequency Spectrum Representation and Fusion}
% Different from attention mechanisms that learn the dependencies between features at various spatial locations, Fourier transform losslessly transforms spatial domain features into the frequency domain

The frequency spectrum representation and fusion module losslessly transform spatial domain features into the frequency domain, obtaining discriminative spectrum features for each modality. The frequency spectrum gives text and image representations a complete view of spatial features and facilitates obtaining informative components and eliminating irrelevant components from a global view. 

% We first take advantage of the fast Fourier transform (FFT) to perform the domain transformation. 
\subsubsection{Spectrum representation} We first transform the spatial features into spectrum features using Discrete Fourier transform (DFT). The spectrum of text features can be obtained as follows:
% \begin{equation}
%     \mathbf{X}^t=\mathcal{F}_{seq}(\mathcal{F}_{h}(\mathbf{x}^t))\in \mathbb{C}^{m\times d}
% \end{equation}
\begin{equation}
    \mathbf{X}^t[k]=\mathcal{F}_{seq}(\mathbf{x}^t[i])=\sum_{i=0}^{m-1}\mathbf{x}^t[i]e^{-j(2\pi /m)ki}  
\end{equation}
where $\mathbf{X}^t\in \mathbb{C}^{m\times d}$ is a complex tensor, $\mathbf{X}^t[k]$ is the spectrum of $\mathbf{x}^t[i]$ at the frequency $2\pi k/m$, $\mathcal{F}_{seq}(\cdot)$ is the 1D DFT along the sequence dimension, and $j$ is the imaginary unit. 
% and $\mathcal{F}_{h}(\cdot)$ denotes the 1D FFT along the hidden dimension. 
The spectrum of image embedding can be obtained:
\begin{equation}
    \mathbf{X}^v[k]=\mathcal{F}_{pat}(\mathbf{x}^v[i])=\sum_{i=0}^{n-1}\mathbf{x}^v[i]e^{-j(2\pi /m)ki}
\end{equation}
where $\mathbf{X}^v\in \mathbb{C}^{n\times d}$ is a complex tensor, $\mathcal{F}_{pat}(\cdot)$ denotes the 1D DFT along the patch dimension. 
Self-attention computes the spatial dependencies in a quadratic time complexity, while DFT can be efficiently implemented via a fast Fourier transform in logarithmic time complexity. Refer to Appendix C\ref{app:anatc} for a more detailed comparison.
% Furthermore, the fine-grained cross-modal spectrum co-selection facilitates the common analysis of spectral components in text and images during the inference process and guarantees the complementation between multimodal rumor features, which allows the model to retain the informative components more properly.

% Note that DFT has conjugate symmetric for real input \cite{RaoZZLZ21}, meaning that the information about the real input frequency characteristics is contained in half of its DFT. The fast Fourier transform (FFT) and inverse fast Fourier transform (IFFT) are able to be utilized to implement the DFT and IDFT algorithms to achieve efficient computation. 

\subsubsection{Unimodal spectrum compression (USC)}
Spatial features are effectively consolidated within each frequency element, enabling the extraction of informative features from both text and images through the point-wise product in the frequency domain. We introduce a filter bank for each modality $\mathbf{X}^a,a\in \{t, v\}$ to compress the spectrum and obtain the significant features associated with rumors. 
We use $\mathbf{K}^a=[\mathbf{k}^a_1, \mathbf{k}^a_2,...,\mathbf{k}^a_k]$ to represent the filter bank, where $k$ is the number of filters in the filter bank:
% The calculation process is as follows:
\begin{equation}
    \hat{\mathbf{X}}^a=\sum_{i=1}^{k}\frac{1}{l}{|\mathbf{X}^a|}^2\odot\mathbf{k}^a_i cos(\frac{(2i-1)\pi}{2k}), a\in\{t,v\}
    \label{eq:filter}
\end{equation}
where $\odot$ is the element-wise multiplication, $|\mathbf{X}^a|^2$ is the power spectrum of $\mathbf{X}^a$, $l$ is the length of $\mathbf{X}^a$. 
The $|\mathbf{X}^a|^2$ operation smooths the spectrum, highlighting the main components of the spectrum from an intra-modal perspective. It also facilitates the subsequent learning of unimodal compression. 
% $cos((2i-1)\pi /2k)$ refers to the Discrete Cosine Transform module, which compacts better energy and can aggregate the more important information in the rumor features. 
$cos((2i-1)\pi /2k)$ compacts better energy and can aggregate the more important information in the rumor features. 
Its combination of application with the filter bank $\mathbf{K}^a$ allows for efficient frequency domain feature compression. 

\subsubsection{Cross-modal spectrum co-selection (CSC)}
% After passing the uni-modal frequency compression, we select the spectrum components from a cross-modal perspective. 
% We postulate that some spectrum components are less beneficial for rumor detection and that removing the rumor-irrelevant spectrum components would benefit detection performance. Accordingly, we propose an emphasize and suppress (E\&S) module to emphasize informative components and suppress irrelevant components for each modality, co-attending to each unimodal spectrum. 
Based on the postulation that certain spectrum components have limited contributions to rumor detection, we propose an emphasize and suppress (E\&S) module, which aims to enhance informative components and suppress irrelevant components within each modality by co-attending to the unimodal spectrum. We first perform average pooling over the compressed spectrum $\hat{\mathbf{X}}^a,a\in \{t,v\}$, subsequently applying convolution to obtain the representation of the rumor visual/text clues. 
Consequently, we can derive two selection filters, one from the visual spectrum and another from the text spectrum. The filters serve the purpose of co-selecting informative features from each other. 
We perform cross-modal spectrum co-selection by multiplying the two filters with the corresponding unimodal spectrum in a staggered manner:
% \begin{equation}
%     \mathbf{\Lambda}^a = \operatorname{Conv}(\operatorname{Avg}(\hat{\mathbf{X}}^a\odot\mathbf{\Theta}^a))
%     \label{eq:gate}
% \end{equation}
\begin{align}
    \tilde{\mathbf{X}}^t = \hat{\mathbf{X}}^t\odot\operatorname{Conv}(\operatorname{Avg}(\hat{\mathbf{X}}^v\odot\mathbf{\Theta}^v)) \\
    \tilde{\mathbf{X}}^v = \hat{\mathbf{X}}^v\odot\operatorname{Conv}(\operatorname{Avg}(\hat{\mathbf{X}}^t\odot\mathbf{\Theta}^t))
    \label{eq:select}
\end{align}
where $\odot$ is the element-wise multiplication, $\mathbf{\Theta}^a$ denotes the trainable parameters with the same dimension as $\hat{\mathbf{X}}^a$, $Conv(\cdot)$ is an $1\times 1$ convolutional layer, and $Avg(\cdot)$ is the average pooling function. The convolutional layer and $\mathbf{\Theta}^a$ facilitate learning how to emphasize informative components and suppress irrelevant components for multimodal fusion.

% \begin{align}
%     \tilde{\mathbf{X}}^t = \hat{\mathbf{X}}^t\odot \mathbf{\Lambda}^v \\
%     \tilde{\mathbf{X}}^v = \hat{\mathbf{X}}^v\odot \mathbf{\Lambda}^t
%     \label{eq:select}
% \end{align}
Finally, we employ inverse discrete Fourier transform (IDFT, $\mathcal{F}^{-1}_{seq}$ and $\mathcal{F}^{-1}_{pat}$) to convert the spectral representations of text and image back into the spatial domain:% and update $\mathbf{x}^a, a\in\{t,v\}$: 
\begin{gather}
    \mathbf{x}^t\leftarrow\mathcal{F}^{-1}_{seq}(\tilde{\mathbf{X}}^t) \\
    \mathbf{x}^v\leftarrow\mathcal{F}^{-1}_{pat}(\tilde{\mathbf{X}}^v)
\end{gather}
The fine-grained cross-modal spectrum co-selection facilitates the common analysis of spectral components in text and images during the inference process and guarantees the fusion of multimodal rumor features, which allows the retention of the informative components more properly.

\subsection{Rumor Detection with Contrastive Learning}
\subsubsection{Contrastive Learning Objectives}
To promote multimodal learning in training, we introduce a dual contrastive learning module, consisting of two parts: 1) fully-supervised intra-modal contrastive learning based on rumor veracity labels $\mathcal{L}_{full}$, and 2) self-supervised inter-modal contrastive learning based on multimodal spatial semantics $\mathcal{L}_{self}$. 
% With the former, the representations of the same label can be placed closer together in the semantic space and those of different labels can be placed farther apart. 
%, making it easier to understand the similarities and differences between classes. 
% With the latter, the exploration of semantic correlations between multimodal features is made possible, and the same sample of text and image can derive more robust multimodal representations. 
% The detailed formulas for both loss functions can be found in the Appendix. 

In a mini-batch $\mathcal{B}$, we divide samples according to the rumor veracity label into $R_0, R_1$. 
For the anchor sample $r_i\in R_1$, the positive pair can be denoted as $(r_i,r_j)$, where $r_j\in R_1, j\neq i$. 
The samples in $R_0$ are regarded as negative examples. 
As such, we follow \cite{LinLLDYZX22} to define the pairwise objective function with anchor sample and positive or negative samples $\mathcal{L}_1(\mathbf{x}^a,\mathbf{x}^a),a\in \{t,v\}$.  
% Then, we calculate the contrastive loss separately for textual and visual modal representations and then sum up to obtain the final fully-supervised intra-modal contrastive loss as follows: 
The final fully-supervised intra-modal contrastive loss is as follows:
\begin{equation}
\begin{aligned}
    \mathcal{L}_{full}=\sum_{\mathcal{M}}[\sum_{r_i\in R_1}\frac{1}{|R_1|}\sum_{j,r_j\in R_1,j\neq i}\mathcal{L}_1(\mathbf{x}^a_i,\mathbf{x}^a_j)+\\
    \sum_{r_k\in R_0}\frac{1}{|R_0|}\sum_{l,r_l\in R_1,l\neq k}\mathcal{L}_1(\mathbf{x}^a_k,\mathbf{x}^a_l)]
\end{aligned}
\end{equation}
where $|\cdot|$ denotes the number of corresponding samples.

% and $|R_1|$ denote the number of samples in $R_0$ and $R_1$ respectively. 

For self-supervised inter-modal contrastive loss, we consider the text and associated image of the given anchor sample $r_i$ to be a positive sample, while the other pairs are considered negative samples.
We use the InfoNCE loss \cite{He0WXG20} to optimize the image and text features, denoted as $\mathcal{L}_2(\mathbf{x}^t,\mathbf{x}^v)$ and $\mathcal{L}_2(\mathbf{x}^v,\mathbf{x}^t)$. 
And the self-supervised inter-modal contrastive loss is as follows:
\begin{gather}
    \mathcal{L}_{self}=\frac{1}{2|\mathcal{B}|}\sum^{|\mathcal{B}|}_{i=1}[\mathcal{L}_2(\mathbf{x}^t_i,\mathbf{x}^v_i)+
    \mathcal{L}_2(\mathbf{x}^v_i,\mathbf{x}^t_i)]
\end{gather}
where $|\mathcal{B}|$ denotes the number of samples in mini-batch $\mathcal{B}$. 

\subsubsection{Detection based on distribution similarity} After obtaining the improved text and image representations, we measure the Jensen-Shannon (JS) divergence between the two features to learn the distribution similarity, which is subsequently utilized to control the final multimodal rumor representation output. 
Since it is difficult to infer the posterior probability $p$ from the given data sample, we generate an approximation of its distribution $q$. 
Specifically, the posterior probability of unimodal can be denoted separately as $q(z^t|x^t)$ and $q(z^v|x^v)$. 
The divergence of different modal distributions in $\mathbf{x}^a$ can then be measured as follows:
% \begin{gather}
%     \gamma^t=\frac{1}{2}\sum^N_{i=1}q(z^t_i|x^t_i)log(\frac{2q(z^t_i|x^t_i)}{q(z^t_i|x^t_i)+q(z^v_i|x^v_i)}) \\
%     \gamma^v=\frac{1}{2}\sum^N_{i=1}q(z^v_i|x^v_i)log(\frac{2q(z^v_i|x^v_i)}{q(z^t_i|x^t_i)+q(z^v_i|x^v_i)}) \\
%     \gamma=\gamma^t+\gamma^v
% \end{gather}
\begin{gather}
    \gamma=\operatorname{JS}(q(z^t|x^t)||q(z^v|x^v))
\end{gather}
where $JS(\cdot)$ denotes the JS divergence, and the similarity score $\gamma$ is computed by the JS divergence. Accordingly, we can calculate the integrated multimodal representation and apply a fully connected layer $\operatorname{FC}$ to predict the label $\hat{y}$: 
\begin{equation}
    \mathbf{m}=(1-\gamma)(\mathbf{W}^t\mathbf{x}_t+\mathbf{W}^v\mathbf{x}_v)+\gamma\mathbf{x}_t+\gamma\mathbf{x}_v
\end{equation}
\begin{equation}
    \hat{y}=\operatorname{Softmax}(\operatorname{FC}(\mathbf{m}))
\end{equation}
where $\mathbf{W}^t$ and $\mathbf{W}^v$ are trainable parameters, and $\gamma$ is a hyperparameter to adaptively weigh cross-modal features.%control the contribution of

Taking rumor detection as a binary classification task, we then apply the cross-entropy loss as the detection objective:
\begin{equation}
    \mathcal{L}_{cls}=-\mathbb{E}_{y\sim\hat{Y}}[ylog(\hat{y})+(1-y)log(1-\hat{y})]
\end{equation}
Finally, the final loss can be written as:
\begin{equation}
    \mathcal{L}=\mathcal{L}_{cls}+\alpha\mathcal{L}_{full}+\beta\mathcal{L}_{self}
\end{equation}
with hyperparameters $\alpha$, $\beta$ to balance different objectives.
% controlling the effect of different losses.

\section{Experiments}

\begin{table*}[!t]
\caption{Performance comparison on the Weibo and Twitter datasets. The best performance is highlighted in bold, while underlining highlights the follow-up, and $*$ indicates the statistically significant improvement (i.e., two-sided $t$-test with $p<0.05$).}
\vspace{-3mm}
\centering
\begin{tabular}{ccccccccc}
\hline
                           &                           &                                       & \multicolumn{3}{c}{Rumor}                                                                                             & \multicolumn{3}{c}{Non-rumor}                                                                                         \\ \cline{4-9} 
\multirow{-2}{*}{Datasets} & \multirow{-2}{*}{Methods} & \multirow{-2}{*}{Accuracy}            & Precision                             & Recall                                & F1                                    & Precision                             & Recall                                & F1                                    \\ \hline
                            & att-RNN \cite{jin2017multimodal}  & 0.772                                 & 0.854                                 & 0.656                                 & 0.742                                 & 0.720                                 & 0.889                                 & 0.795                                 \\
                           & EANN \cite{WangMJYXJSG18}  & 0.827                                 & 0.847                                 & 0.812                                 & 0.829                                 & 0.807                                 & 0.843                                 & 0.825                                 \\
                           & MVAE \cite{KhattarG0V19}  & 0.824                                 & 0.854                                 & 0.769                                 & 0.809                                 & 0.802                                 & 0.875                                 & 0.837                                 \\
                           & SpotFake \cite{SinghalS0KS19}  & \underline{0.892}                                 & 0.902                                 & \textbf{0.964}                                 & \textbf{0.932}                                 & 0.847                                 & 0.656                    & 0.739                  \\
                           & HMCAN \cite{QianWHFX21}   & 0.885                                 & \underline{0.920} & 0.845                                 & 0.881                                 & 0.856                                 & \textbf{0.926} & \underline{0.890}                                  \\
                           & CAFE \cite{LZSLTS22}      & 0.840                                 & 0.855                                 & 0.830                                 & 0.842                                 & 0.825                                 & 0.851                                 & 0.837                                 \\
                           & BMR \cite{ying2023bootstrapping} & 0.884 & 0.875                                 & 0.886 &  0.880 & \underline{0.874} & 0.881                                 & 0.877 \\
                           & LogicDM \cite{LiuWL23}    & 0.852                                 & 0.862                                 & 0.845                                 & 0.853                                 & 0.843                                 & 0.859                                 & 0.851                                 \\
\multirow{-8}{*}{Weibo}    & \textbf{FSRU}             & \textbf{0.901}* & \textbf{0.922}* & \underline{0.892} & \underline{0.906} & \textbf{0.879}* &  \underline{0.913} & \textbf{0.895}* \\ \hline
                           & att-RNN \cite{jin2017multimodal}  & 0.664                                 & 0.749                                 & 0.615                                 & 0.676                                 & 0.589                                 & 0.728                                 & 0.651                                 \\
                            & EANN \cite{WangMJYXJSG18}  & 0.648                                 & 0.810                                 & 0.498                                 & 0.617                                 & 0.584                                 & 0.759                                 & 0.660                                 \\
                           & MVAE \cite{KhattarG0V19}  & 0.745                                 & 0.801                                 & 0.719                                 & 0.758                                 & 0.689                                 & 0.777                                 & 0.730                                 \\
                           & SpotFake \cite{SinghalS0KS19}  & 0.777                                 & 0.751                                 & \underline{0.900}                                 & 0.820                                 & 0.832                                 & 0.606                                 & 0.701                                 \\
                           & HMCAN \cite{QianWHFX21}   & 0.897                                 & \underline{0.971} & 0.801                                 & \underline{0.878} & 0.853                                 & \underline{0.979} & 0.912                                 \\
                           & CAFE \cite{LZSLTS22}      & 0.806                                 & 0.807                                 & 0.799                                 & 0.803                                 & 0.805                                 & 0.813                                 & 0.809                                 \\
                           & BMR \cite{ying2023bootstrapping} & 0.872                                 & 0.842                                 & 0.751                                 & 0.794                                 & 0.885                                 & 0.931                                 & 0.907                                 \\
                           & LogicDM \cite{LiuWL23}    & \underline{0.911} & 0.909                                 & 0.816                                & 0.859                                 & \textbf{0.913} & 0.958 & \underline{0.935} \\
\multirow{-7}{*}{Twitter}  & \textbf{FSRU}             & \textbf{0.952}* & \textbf{0.983}* & \textbf{0.938}* & \textbf{0.960}* & \underline{0.901} & \textbf{0.984}* & \textbf{0.940}* \\ \hline
\end{tabular}
\label{table:baselines}
\vspace{-2mm}
\end{table*}

In this section, we evaluate the effectiveness of our proposed model \footnote{https://github.com/dm4m/FSRU} on two real-world datasets. 

\subsection{Experimental Setup}
\subsubsection{Datasets}
To facilitate comparison with the baselines, we evaluate the proposed FSRU on two publicly available multimodal datasets: Twitter \cite{boididou2014challenges} and Weibo \cite{jin2017multimodal}. We comprehensively describe each dataset in Appendix B.1\ref{app:datasets}. 

\subsubsection{Baselines}
We compare our FSRU to recent baseline models:
att-RNN \cite{jin2017multimodal}, EANN \cite{WangMJYXJSG18}, MVAE \cite{KhattarG0V19}, SpotFake \cite{SinghalS0KS19}, HCMAN \cite{QianWHFX21}, CAFE \cite{LZSLTS22}, BMR \cite{ying2023bootstrapping}, and LogicDM \cite{LiuWL23}. 
We comprehensively describe each baseline in Appendix B.2\ref{app:baselines} and explain the rationale behind selecting these specific baselines.

\subsubsection{Settings}
We implemented our algorithms using PyTorch 1.12 and conducted all experiments on a single NVIDIA RTX 3080 Ti GPU. 
The loss function is optimized using the Adam algorithm \cite{kingma2014adam}. 
The evaluation metrics include Accuracy, Precision, Recall, and F1 score. 
To ensure fairness, we employ five-fold cross-validation for the experiments. 
We utilize publicly available Word2Vec \cite{corr/abs-1301-3781} to obtain the word embeddings. 
Images are resized into 224×224. 
The maximum sequence length is set to 50 for Weibo and 32 for Twitter. 
The dimension of text and image embedding is set to 256. 
The model is trained for 50 epochs with a batch size of 64. 
For Weibo, the initial learning rate is set to 1e-2, while for Twitter, it is set to 1e-5.
When selecting hyper-parameters $\alpha$ and $\beta$, we consider values from the set $\{0.0,0.05,0.1,0.2,0.3,0.4,0.5\}$.
Ultimately, we set $\alpha$ and $\beta$ to 0.2 for both datasets.
The number of filters in unimodal spectrum compression denoted as $k$ is chosen from the set $\{1,2,4,8\}$, and the final value selected for the results is $k=2$.  
To efficiently implement the DFT and IDFT, we utilized the Fast Fourier Transform (FFT) and inverse FFT. 
The code and implementation details can be found in the supplementary materials.

\subsection{Results and Analysis}
The performance comparison between FSRU and eight other baselines on the two datasets is presented in Table \ref{table:baselines}.  
We further investigate the complexity of FSRU in terms of FLOPs and parameter volumes, compared with state-of-the-art methods. The results are shown in Appendix C\ref{app:anatc}. 

Att-RNN, EANN, and MVAE overlook the deep semantic relationships and interactions among features, leading to limitations in their detection accuracy. 
SpotFake leverages pre-trained models to extract text and image features, demonstrating strong performance in classifying rumors but relatively weaker performance in classifying non-rumors. 
The Transformer is utilized as a feature encoder in HMCAN, enabling effective token mixing through self-attention in the spatial domain and facilitating the acquisition of multimodal representations.
% As we previously examined, the interaction between modalities across the semantic space is not lossless. Therefore, the performance of these two methods on both datasets leaves something to be desired.
To effectively aggregate unimodal representations and cross-modal correlations, CAFE utilizes cross-modal alignment and disambiguation mechanisms. While it demonstrates good performance on the Weibo dataset, its effectiveness diminishes when applied to the Twitter dataset. 
BMR leverages multi-view learning to estimate the importance of different modalities for adaptive aggregated unimodal representation, resulting in superior performance. 
LogicDM considers logical relationships between predicates and selects predicates and cross-modal objects to derive and evaluate interpretable logical clauses, resulting in improved performance on the Twitter dataset.

Our proposed FSRU has delivered highly favorable results on both datasets, consistently ranking 1st or 2nd across all evaluation metrics. FSRU effectively explores and integrates multimodal features within the frequency domain. By leveraging the Fourier transform to bridge the spatial and frequency domains, FSRU achieves a lossless transformation of multimodal rumor features into a shared space. 
FSRU takes a cross-modal perspective to control spectral components while also capturing the intrinsic characteristics of rumors from an unimodal perspective. 
This conceptually straightforward yet computationally efficient approach significantly enhances the performance of rumor detection. 
In addition, FSRU employs multimodal feature aggregation based on distributional similarity and two types of contrastive learning to learn the complementary relationships between cross-modal features. This allows FSRU to adaptively aggregate multimodal features for detection. 
However, it is important to note that the impact on the Weibo dataset appears to be slightly less pronounced compared to the Twitter dataset, possibly due to inherent differences between the two datasets. Firstly, the Weibo dataset is relatively smaller in size when compared to the Twitter dataset. Secondly, the Weibo dataset comprises a subset of images that exhibit lower quality or contain less informational content.
% We observe a significant improvement on Twitter, which is probably due to the fact that the Twitter dataset contains multiple texts that correspond to the same image, and the model iteratively mines the images to gather potential features. 

\subsection{Ablation Study}
To assess the effectiveness of different modules within FSRU, we conduct a comparative analysis with sub-models denoted as ``-w/o USC'', ``-w/o CSC'', ``-w/o DSF'', and ``-w/o CL''. 
These variants represent FSRU without considering unimodal spectrum compression, cross-modal spectrum co-selection, distribution similarity-based fusion, and dual contrastive learning, respectively.  
The results are shown in Table \ref{table:ablation} and Figure \ref{fig:tSNE}. 

\begin{table}[!h]
\caption{Comparison of different FSRU variants.}
\vspace{-3mm}
\centering
\begin{tabular}{ccccc}
\hline
\multicolumn{1}{l}{} & \multicolumn{2}{c}{Weibo}         & \multicolumn{2}{c}{Twitter}     \\ \cline{2-5} 
\multicolumn{1}{l}{} & Accuracy        & F1              & Accuracy        & F1              \\ \hline
FSRU                 & \textbf{0.901} & \textbf{0.902} & \textbf{0.952} & \textbf{0.950} \\ \hline
-w/o USC              & 0.866       & 0.865       & 0.910     & 0.908     \\
-w/o CSC              & 0.883       & 0.882       & 0.924     & 0.922     \\ 
-w/o DSF              & 0.876       & 0.875       & 0.947     & 0.943     \\ 
-w/o CL               & 0.889       & 0.889       & 0.937     & 0.936     \\ \hline
% -w/o $\mathbf{x}^v$   & 0.872       & 0.872       & 0.931     & 0.928     \\
% -w/o $\mathbf{x}^t$   & 0.882       & 0.882       &      &      \\ \hline
\end{tabular}
\vspace{-4mm}
\label{table:ablation}
\end{table}

\subsubsection{Quantitative analysis}
As shown in Table \ref{table:ablation}, It is evident that removing either the unimodal spectrum compression or the cross-modal spectrum co-selection adversely affects the model's performance on both datasets.  
% It indicates that modeling multimodal features in the frequency domain from unimodal and cross-modal viewpoints is required. 
Without employing unimodal spectrum compression, the model loses the ability to explore distinctive patterns in modal frequency responses. Similarly, the absence of cross-modal spectrum component interactions hinders the model's capacity to learn dependencies between multimodal features. 
Moreover, excluding the distribution similarity-based fusion and the dual contrastive learning module from the model leads to a slight decline in performance.
These findings highlight the significance of fusing multimodal features by measuring multimodal distribution similarity and leveraging dual contrastive learning. 
% Furthermore, the model's performance experiences a slight decline when the distribution similarity-based fusion module is excluded. 
% This implies that fusing multimodal features based on measuring multimodal distribution similarity holds significance and contributes to improved performance. 
% From the ablation experiments in the dual contrastive learning module, we can see that omitting contrastive learning results in some model performance degradation on both datasets. 

\subsubsection{Qualitative analysis}
\begin{figure}[!h]
\centering
\subfigure{
\includegraphics[width=0.46\textwidth]{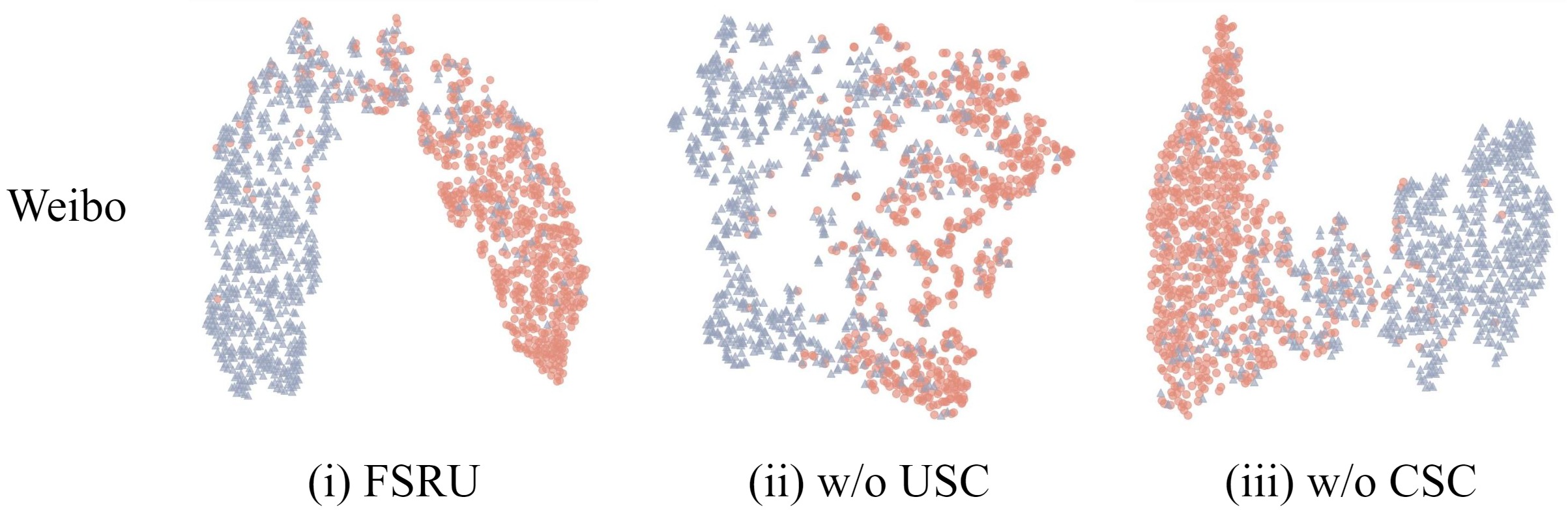} 
}
\subfigure{
\includegraphics[width=0.46\textwidth]{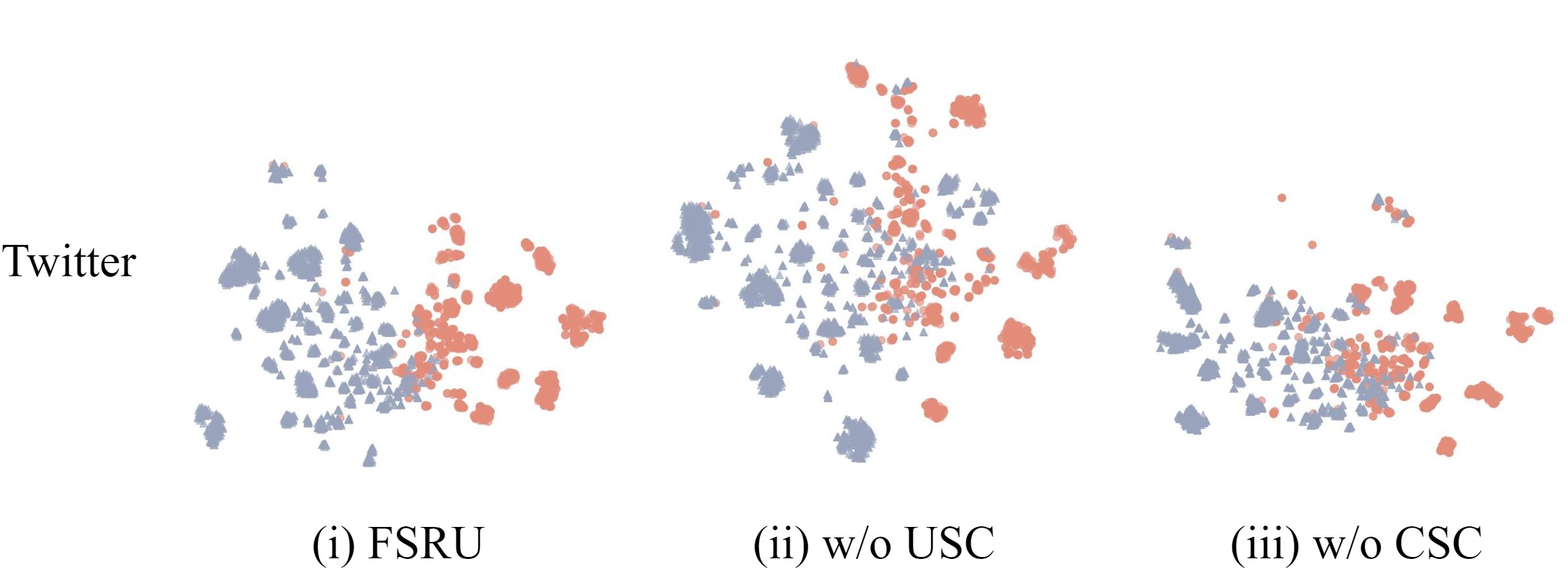} 
}
\DeclareGraphicsExtensions.
\caption{T-SNE visualization of learned representations.}% on two datasets
\label{fig:tSNE}
\vspace{-4mm}
\end{figure}

To further analyze the effect of the frequency spectrum representation and fusion module, we qualitatively visualize the features on the Weibo and Twitter test set with t-SNE \cite{van2008visualizing} as depicted in Figure \ref{fig:tSNE}. 
The FSRU variants ``-w/o USC'' and ``-w/o CSC'' demonstrate the ability to discriminate multimodal rumor features, but there is a clear overlap between features across different labels. 
In contrast, the features learned by FSRU exhibit clear boundaries between labels, effectively reducing the overlapping between features. 
% The comparison between different variants of FSRU proves that the proposed approach learns better multimodal rumor representations, and thus achieves better performance. 

\begin{figure*}[!t]
    \centering
    \includegraphics[width=0.9\textwidth]{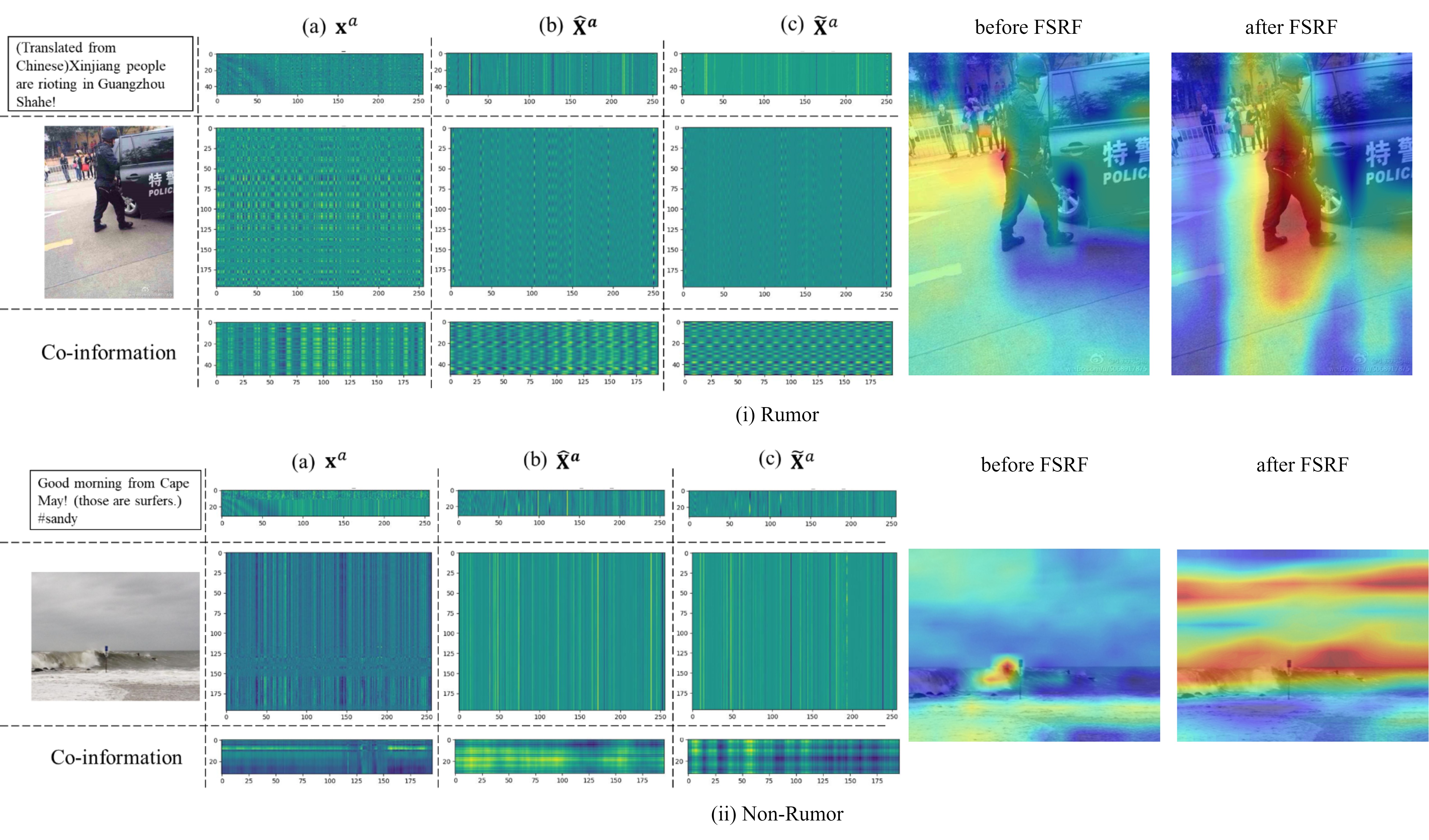}
    \vspace{-3mm}
    \caption{Interpretative visualization of rumor and non-rumor cases. Refer to Appendix D~\ref{app:case} for more illustrative cases.}
    \label{fig:visualization}
    \vspace{-4mm}
\end{figure*}

\subsection{Impact of the Number of Filters $k$}
We conducted experiments by varying the value of $k$ in USC from 1 to 8, as presented in Table \ref{table:filter}. 
The results exhibit a pattern of initially increasing performance followed by a subsequent decline on both datasets. 
Specifically, there is a significant performance improvement from $k=1$ to $k=2$, while a slight decrease is observed from $k=2$ to $k=8$.
% Only single frequency response can be learned by the model when $k=1$. 
% The computational cost of the model increases when $k>2$. 
% Since redundant frequency responses may be learned, the performance of detection is affected. 
By setting $k=2$, the model has the ability to acquire diverse and distinct feature patterns from various dimensions of the frequency response while still maintaining an appropriate computational cost. 
Therefore, we determine that $k=2$ is the optimal choice for FSRU on both datasets. 

\begin{table}[!h]
\caption{Effect of the number of filters in USC.}
\vspace{-3mm}
\centering
\begin{tabular}{ccccc}
\hline
\multirow{2}{*}{Filter} & \multicolumn{2}{c}{Weibo}         & \multicolumn{2}{c}{Twitter}     \\ \cline{2-5} 
                        & Accuracy        & F1              & Accuracy        & F1              \\ \hline
1                       & 0.839           & 0.838          & 0.938      &  0.936   \\ \hline
2                       & \textbf{0.901} & \textbf{0.902} & \textbf{0.952} & \textbf{0.950} \\ \hline
4                       & 0.896           & 0.895          & 0.944      & 0.942     \\ \hline
8                       & 0.894           & 0.893          & 0.931      & 0.928     \\ \hline
\end{tabular}
\vspace{-4mm}
\label{table:filter}
\end{table}

\subsection{Case Study}
To provide an intuitive demonstration of the learning process of the Frequency Spectrum Representation and Fusion (FSRF) in FSRU, we visualize $\mathbf{x}^a$, $\hat{\mathbf{X}}^a$, and $\tilde{\mathbf{X}}^a$ $(a\in{t,v})$, along with the corresponding co-information for the two modalities, as shown in Figure \ref{fig:visualization}.
In the case of rumor, as FSRF is learned, the features gradually acquire a distinct pattern, allowing for better differentiation. 
This results in a clearer identification of concentrated spectral energy.
On the other hand, in the case of non-rumors, the model seeks to capture truthfulness clues expressed through multimodal features to the best of its ability. 
FSRF leverages co-selection across modalities to emphasize and suppress specific spectral features across modalities, thereby potentially revealing cues that indicate the veracity of rumors. 

We have visualized the multimodal features of the two mentioned cases before and after the learning process of FSRF. 
In the first image, the model after FSRF learning concentrates on the person in the image, who does not match the person or event mentioned in the text. However, this person does not correspond to the individual or event mentioned in the accompanying text. This image therefore is classified as a rumor.
In the second image, the model concentrates on the waves, the cloudy sky, and the surfer in the distance. This alignment between the visual elements and the textual description suggests consistency and coherence. Hence this image is classified as a non-rumor.

\section{Conclusion}
We first attempt to introduce a frequency spectrum representation and fusion network (FSRU) for multimodal rumor detection. 
% FSRU is unique with a frequency spectrum representation and fusion to capture and aggregate cross-modal features precisely. 
% In this module, we analyze the frequency response in the frequency domain which illustrates a more distinctive pattern of unimodal features and leverages cross-modal complementary interactions to select informative spectrum components. 
FSRU is unique with a frequency spectrum representation and fusion to effectively capture both the frequency of feature changes and their intensity in the frequency domain, which is essential for FSRU to learn multimodal features properly. 
% Subsequently, we present a distribution similarity-based feature fusion module to adaptively integrate the acquired modal representations and perform the final detection. 
Substantial experiments demonstrate that our proposed approach achieves advanced performance. 
% Our future studies include exploring deep insights and mechanisms in frequency-based multimodal fusion and investigating practical approaches to extract multimodal clues of news veracity to improve multimodal rumor detection. 
% Exploring the concept of early fusion in architecture represents a promising research direction.
% We are committed to exploring the implementation of early convergence within our architecture. 
Our future studies include exploring deep insights and mechanisms in frequency-based multimodal fusion to improve multimodal rumor detection.
The proposed model has the potential for more multimodal tasks and scenarios, we will further investigate the effectiveness and interpretability of the spectrum in multimodal fusion.

\section{Acknowledgments}
The research reported in this study is supported by the National Natural Science Foundation of China (No. 62372043). This work is also supported by the BIT Research and Innovation Promoting Project (Grant No. 2023YCXY037) and the National Key Research and Development Program of China (2022YFB3104702).

\bibliography{aaai24}

\clearpage
\appendix
\section{A. Theoretically Analysis}
In this section, we theoretically analyze the equivalence between self-attention and frequency-domain computation, i.e., we can efficiently reformulate self-attention via point-wise computation in the frequency domain.

Given the input tensor, $X$ we denote the $n$-th token as $x_n\in\mathbb{R}^d$ and define $N$ as the sequence length.
\subsection{Definition 1 (Self-Attention)}
We express the self-attention $\operatorname{Self-Att:\mathbb{R}^{N\times d}\rightarrow\mathbb{R}^{N\times d}}$ using the formulation of kernel integration \cite{tsai2019transformer, cao2021choose, kovachki2021neural, AFNO}:
\begin{equation}
    \operatorname{Self-Att}=\operatorname{softmax}(\frac{XW_q(XW_k)^\top}{\sqrt{d}})XW_v
\end{equation}
Define $K=\operatorname{softmax}((XW_q(XW_k)^\top)/\sqrt{d})$as the $N\times N$ score array. Then the self-attention can be treated as an asymmetric matrix-valued kernel $\kappa=[N]\times[N]\rightarrow\mathbb{R}^{d\times d}$ parameterized as $\kappa[s,t]=K[s,t]\circ W_v^\top$. Therefore, self-attention can be viewed as a kernel summation.
\begin{equation}
    \operatorname{Self-Att}(X)[s]=\sum_{t=1}^N\kappa[s,t]X[t]\quad\forall s\in[N]
\end{equation}
The concept of kernel summation can be extended to encompass continuous kernel integrals.  The input tensor $X$ represents a spatial function in the function space $X\in(D,\mathbb{R}^d)$, where it is defined on a domain $D$:
\begin{equation}
    \operatorname{Self-Att}(X)[s]=\mathcal{K}(X)(s)=\int_D\kappa(s,t)X(t)\,\text{d}t\quad\forall s\in D
\end{equation}
where for the continuous input $X\in D$, the kernel integral $\mathcal{K}:(D,\mathbb{R}^d)\rightarrow(D,\mathbb{R}^d)$ is defined as \cite{AFNO}.

\subsection{Definition 2 (Global Convolution)}
Assuming a green kernel $\kappa(s, t)=\kappa(s-t)$, the above kernel integral leads to global convolution:
\begin{equation}
    \mathcal{K}(X)(s)=\int_D\kappa(s-t)X(t)s\,\text{d}t\quad\forall s\in D
\end{equation}
The convolution is a smaller complexity class of operation compared to integration. Furthermore, the global convolution can be efficiently implemented by the fast Fourier transform in the frequency domain.

\subsection{Frequency-Domain Computation}
As per the convolution theorem \cite{soliman1990continuous}, global convolution in the spatial domain can be equivalently represented as multiplication in the frequency domain. Therefore, for the continuous input $X\in D$ the kernel integral \cite{AFNO} is defined as:
\begin{equation}
    \mathcal{K}(X)(s)=\mathcal{F}^{-1}(\mathcal{F}(\kappa)\cdot\mathcal{F}(X))(s)\quad\forall s\in D
\end{equation}
where $\cdot$ is the point-wise multiplication and $\mathcal{F},\mathcal{F}^{-1}$ is the continuous Fourier transform and inverse Fourier transform.

In summary, employing frequency-domain computation to reformulate self-attention is an efficient and theoretically-equivalent alternative. This analysis further theoretically guarantees the reasonableness and feasibility of our proposed method: using the frequency spectrum to represent and fuse multimodal data.

\section{B. Experimental Details}
\subsection{B.1 Datasets}
\label{app:datasets}
In order to facilitate comparison with the baselines, we evaluate the proposed frequency spectrum representation and fusion network on two publicly available multimodal datasets:
\begin{itemize}
    \item The Twitter dataset \cite{boididou2014challenges}: collected from Twitter and released for Twitter Verifying Multimedia Use task. The training set contains 4,992 real tweets and 9,470 rumor tweets. The testing set contains 1,215 real tweets and 717 rumor tweets.%\footnote{https://www.twitter.com}
    \item The Weibo dataset \cite{jin2017multimodal}: collected from XinHua News Agency and Weibo. The training set contains 3,783 real tweets and 3,749 rumor tweets. The testing set contains 996 real tweets and 1,000 rumor tweets. %\footnote{http://www.xinhuanet.com/}\footnote{https://weibo.com}
\end{itemize}
Following~\cite{SunZML21, LZSLTS22}, we remove those instances without any text or image since the goal is to perform multimodal rumor detection by fusing text and image information. In addition, if a tweet has more than one corresponding image, we will choose one at random. 

% !!!!!!!!!!!!!

\subsection{B.2 Baselines}
\label{app:baselines}
We compare our proposed model with several state-of-the-art baselines listed as follows: 
\begin{itemize}[leftmargin=*]
    \item att-RNN \cite{jin2017multimodal}: att-RNN uses a recurrent neural network with an attention mechanism to extract multimodal features and to learn the relationships between visual features and joint text/social features. 
    \item EANN \cite{WangMJYXJSG18}: EANN utilizes an adversarial network to improve the fake news detection performance. It consists of three components: the multi-modal feature extractor, the fake news detector, and the event discriminator.
    \item MVAE \cite{KhattarG0V19}: MVAE employs a multimodal variational autoencoder to reconstruct the two modalities from the learned shared representation, and thus discovers the cross-modality association.
    \item SpotFake \cite{SinghalS0KS19}: SpotFake uses BERT to fuse contextual features and uses VGG-19 to learn the image features. Then, for the detection, the two modal representations are joined. 
    \item HCMAN \cite{QianWHFX21}: HCMAN leverages BERT and ResNet to obtain representations for text and image respectively and models the multi-modal context information and the hierarchical semantics of text jointly in a unified deep model.     
    \item CAFE \cite{LZSLTS22}: CAFE can adaptively aggregate discriminative cross-modal correlation features and unimodal features based on the inherent cross-modal ambiguity. 
    \item BMR \cite{ying2023bootstrapping}: BMR proposes the Improved Multi-gate Mixture-of-Expert networks (iMMoE), which learn information from unimodal and multimodal features through single-view prediction and cross-modal consistency learning.
    \item LogicDM \cite{LiuWL23}: LogicDM introduces five meta-predicates and integrates interpretable logic clauses to express the reasoning process of the target task. 
\end{itemize}

\section{C. Analysis of Complexity}
\label{app:anatc}
We conducted a comparison between FSRU and three baseline models, namely BMR, CAFE, and SpotFake, in terms of FLOPs and parameters. As shown in Table \ref{table:complexity}, the proposed FSRU outperforms BMR and SpotFake while requiring lower computational complexity. CAFE demonstrates the lowest computational complexity among the considered models. However, due to relying solely on encoders and MLPs, CAFE falls short in detection performance compared with SOTA baselines.

\begin{table}[!h]
\centering
\caption{Comparison of trainable parameters and computational speed. * indicates results from baseline papers.}
\label{table:complexity}
\begin{tabular}{c|c|c|c|c}
\hline
      & FSRU  & BMR*   & CAFE* & SpotFake* \\ \hline
Param & 1.13M & 94.39M & 0.68M & 124.37M \\ \hline
FLOPs & 9.05G & 18.42G & 0.01G & 30.42G \\ \hline
\end{tabular}
\end{table}

We also compare our frequency spectrum representation and fusion module with the core module/operator (i.e., Spatial MLP and self-attention) for representing/fusing multimodal data in recent prevalent baselines. The results, presented in Table \ref{table:flops}, demonstrate the superior effectiveness of our proposed module over both approaches.

\begin{table}[htbp]
\caption{Complexity of Spatial MLP, Self-Attention, and our proposed frequency spectrum representation and fusion module. $n:=hw$, $m$, and $d$ refer to the sequence size for the image, the sequence size for the text, and the dimensionality, respectively.}
\label{table:flops}
\begin{tabular}{c|cc}
\hline
\multirow{2}{*}{Models} & \multicolumn{2}{c}{Complexity (FLOPs)} \\ \cline{2-3} 
                        & \multicolumn{1}{c|}{image}    & text   \\ \hline
Spatial MLP             & \multicolumn{1}{c|}{$n^2d$}        & $m^2d$     \\ \hline 
Self-Attention          & \multicolumn{1}{c|}{$nd^2+n^2d$}  & $m^2d+md^2$     \\ \hline 
Ours module             & \multicolumn{1}{c|}{$nd\text{log}(n)+(n+d)d$} & \makecell{$md\text{log}(m)+$ \\ $(m\text{log}(d)+d)d$}     \\ \hline
\end{tabular}
\vspace{-2mm}
\end{table}

\section{D. Training Convergence}
\begin{figure}[!h]
    \centering
    \includegraphics[width=0.47\textwidth]{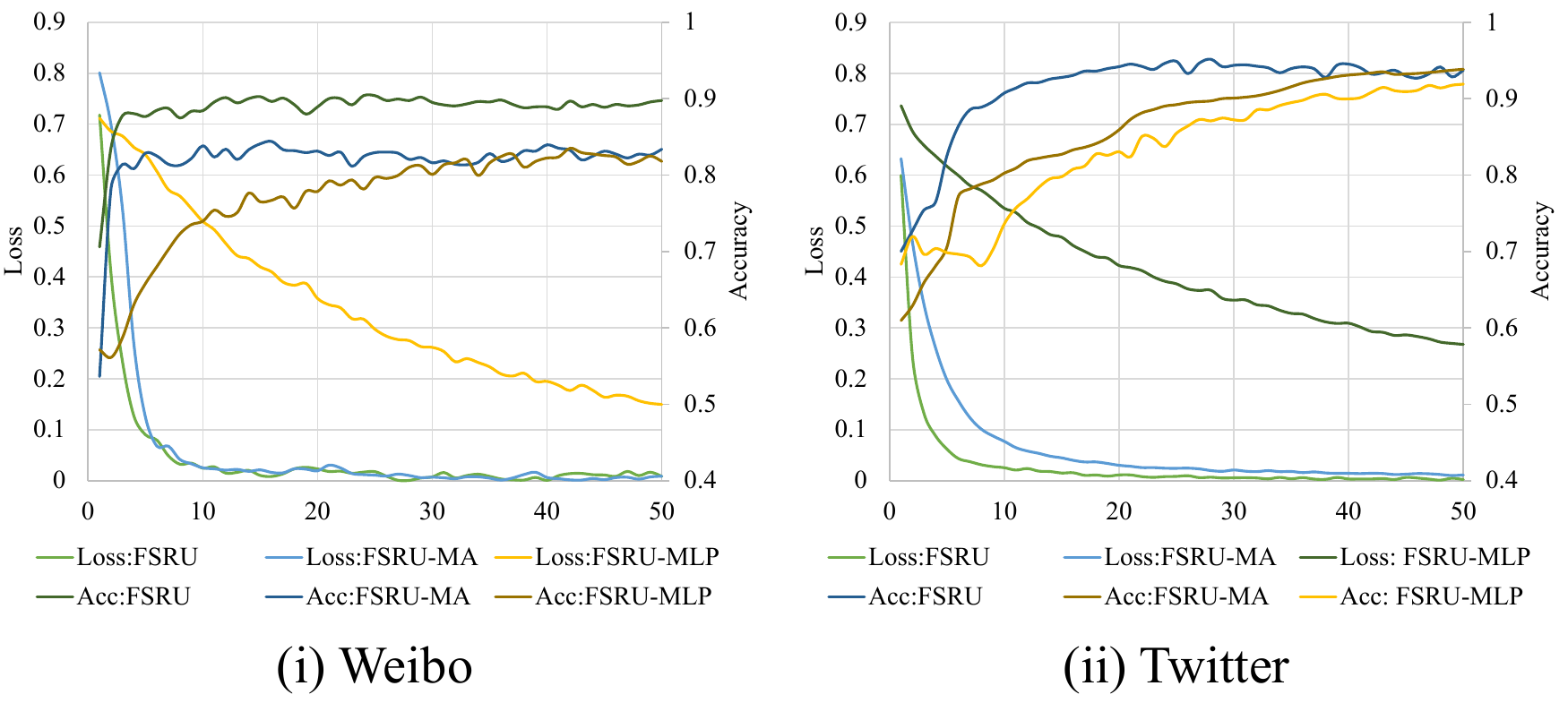}
    \caption{Training loss curve and testing accuracy curve for FSRU and FSRU-MA. The x-axis denotes training epochs.}
    \label{fig:convergence}
    \vspace{-4mm}
\end{figure}

To further validate the convergence performance of the frequency spectrum representation and fusion module in FSRU, we conduct a comparison with the multi-head attention and spatial-MLP methods. 
In particular, we replace the frequency domain functions with the multi-head attention and spatial-MLP within the frequency spectrum representation and fusion module, resulting in a variant denoted as FSRU-MA and FSRU-MLP, respectively. 
In Figure \ref{fig:convergence}, we present a comparison of the loss and accuracy performance separately on both datasets. 
It reports that FSRU converges faster and achieves better detection results than FSRU-MA, indicating the efficiency and effectiveness advantages of our spectrum representation and fusion over self-attention. 
We also observed that the spatial-MLP-based model exhibits inferior classification and convergence performance, despite its advantages of lower computational complexity and shorter training time.
% On Weibo, the accuracy of FSRU is significantly higher than that of FSRU-MA. 
% And the performance of FSRU gradually stabilizes after about 8 iterations whereas FSRU-MA requires 10 iterations to achieve stable performance. 
% On Twitter, the performance of FSRU gradually stabilizes after approximately 10 iterations, whereas FSRU-MA requires approximately 20 iterations to achieve stable performance. 
% its accuracy does approach FSRU until after approximately 40 iterations. 

\begin{figure*}[!h]
    \centering
    \includegraphics[width=0.94\textwidth]{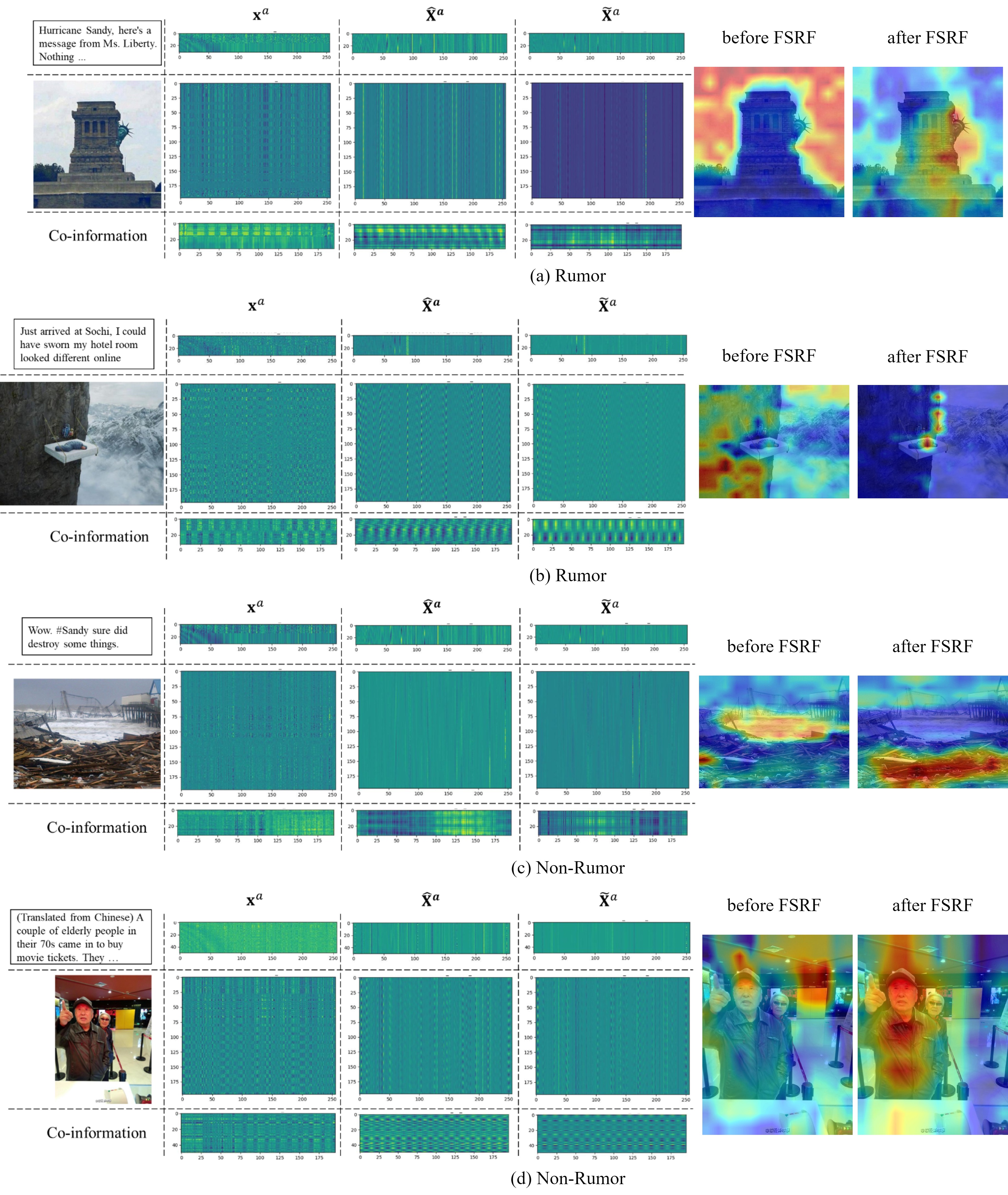}
    \caption{More visualization cases.}
    \label{fig:more}
    \vspace{-3mm}
\end{figure*}

\section{E. More Case Study}
\label{app:case}
In this section, we provide additional visualization cases of both rumors and non-rumors, as shown in Figure \ref{fig:more}. 

Initially, we analyze the transformations in the multimodal features of each example by comparing their states before and after the spectral analysis. 
(1) In Figure \ref{fig:more}.(a), our model, when combined with the accompanying text, identifies the presence of the Statue of Liberty in the image. However, the presence of Lady Liberty in this context is illogical. Upon closer examination, it becomes apparent that the image has been post-processed or manipulated, indicating that the corresponding tweet is a rumor. 
(2) In Figure \ref{fig:more}.(b), by considering the textual cues, the model directs its attention towards the person lying on the mattress and the edge of the crag depicted in the figure. However, the presence of these elements does not align with common-sense expectations. As a result, the model classifies this example as a rumor. 
(3) In Figure \ref{fig:more}.(c), following the analysis using FSRF, the model successfully classified the tweet as a non-rumor by considering the textual content, particularly the phrase "destroy some things," in conjunction with the presence of floating wood depicted in the accompanying picture. 
(4) In Figure \ref{fig:more}.(d), in this scenario, the text depicts two elderly individuals prepared to purchase movie tickets, which aligns with the description provided in the accompanying image. The model correctly localizes the elderly and accurately classifies the corresponding tweet as a non-rumor.

Overall, we can observe that non-rumors tend to exhibit a broader focus on spectral features as they are typically grounded in factual information, resulting in more consistent textual and visual descriptions. Consequently, the frequency spectrum analysis captures various plausible aspects embedded within the multimodal states.
Conversely, rumors are often built on fabricated facts and manipulated images. In such cases, frequency spectrum analysis serves as a means to detect crucial traces of forgery. As a result, the spectral features associated with rumors tend to be concentrated within specific ranges of hidden states, indicating the presence of anomalies or inconsistencies.

In summary, as aforementioned, the Fourier transform offers a sparse frequency spectrum representation for multimodal features, in contrast to the initial embedding. This spectral representation transcends the limitations of location perception and enables the discovery of informative hidden states within each modality from a global view, leading to a more comprehensive learning of the intricate location dependencies present in multimodal features. Hence, we argue utilizing frequency spectrum analysis benefits more effective and interpretable multimodal rumor detection.

\end{document}